\documentstyle[11pt,newpasp,twoside,epsf]{article}
\def\simgt{\lower 2pt \hbox{$\, \buildrel {\scriptstyle >}\over {\scriptstyle \sim}\,$}}
\def\simlt{\lower 2pt \hbox{$\, \buildrel {\scriptstyle <}\over {\scriptstyle \sim}\,$}}
\def\asca{{\it ASCA\/}}
\def\astroe2{{\it Astro-E2\/}}
\def\chandra{{\it Chandra\/}}
\def\conx{{\it Constellation-X\/}}

\def\genx{{\it Generation-X\/}}
\def\hst{{\it {\it HST}\/}}
\def\rosat{{\it ROSAT\/}}
\def\sax{{\it BeppoSAX\/}}
\def\swift{{\it Swift\/}}
\def\xeus{{\it XEUS\/}}
\def\xmm{{\it XMM-Newton\/}}
\def\aox{{$\alpha_{\rm ox}$}}

\reversemarginpar

\begin{document}
\title{X-raying Active Galaxies Found and Missed by the Sloan Digital
Sky Survey}
\author{W.N. Brandt, D.P. Schneider}
\affil{Department of Astronomy \& Astrophysics, The Pennsylvania 
State University, 525 Davey Lab, University Park, PA 16802, USA}
\author{C. Vignali}
\affil{INAF-Osservatorio Astronomico di Bologna, Via Ranzani, 1, 
40127 Bologna, Italy}


\begin{abstract}
Current X-ray observatories, archival X-ray data, and the Sloan Digital 
Sky Survey (SDSS) represent a powerful combination for addressing 
key questions about active galactic nuclei (AGN). We describe a few 
selected issues at the forefront of X-ray AGN research and the 
relevance of the SDSS to them. 
Bulk X-ray/SDSS AGN investigations, 
X-ray weak AGN, 
red AGN, 
hard X-ray selected AGN, 
high-redshift AGN demography, and
future prospects
are all briefly discussed. 
\end{abstract}


\section{Introduction}

X-ray emission appears to be a universal property of AGN, and many 
AGN emit a significant fraction of their total power in 
the X-ray band. 
As a result, X-ray surveys have proved powerful in finding AGN; 
they minimize absorption biases and dilution by host-galaxy light. 
X-ray surveys have found the highest known sky densities of AGN; in 
the \chandra\ Deep Field-North (CDF-N) and South (CDF-S) the AGN density
is $\simgt 5000$~deg$^{-2}$, about an order of magnitude higher than
for the deepest small-area AGN surveys in the optical. 

Detailed X-ray spectral and variability studies probe the immediate vicinity
of the central black hole, where accretion and black hole growth occur, as 
well as the larger scale nuclear environment. 
Primary X-ray continuum components include a hard power law, a soft 
X-ray excess, and a ``reflection'' continuum; these are thought to be 
collectively generated by the inner accretion disk (within \hbox{$\approx 10$--100} 
Schwarzschild radii) and a hot disk corona, and they can show rapid and
large-amplitude variability on timescales down to $\approx 100$--1000~s.  
Radio-loud AGN additionally show power-law emission associated with jet-linked 
X-rays. Many atomic spectral features are also observed including lines and 
edges from ionized outflows and a fluorescent iron K$\alpha$ line associated
with the ``reflection'' process (see Fig.~1 for an example). 


Current X-ray observatories (e.g., \chandra\ and \xmm), 
archival X-ray data (e.g., the thousands of observations made by
\asca, \sax, and \rosat), and 
the SDSS represent a powerful combination for addressing key 
questions about AGN. The SDSS has already demonstrated its power
as an \hbox{X-ray} source identification ``machine'' and has also generated 
large and well-defined AGN samples for X-ray follow-up studies. Below 
we will describe a few selected issues at the forefront of X-ray AGN 
research and the relevance of the SDSS to them. Only limited citations 
will be possible due to finite space; our apologies in advance.

\begin{figure}[t!]
\plotone{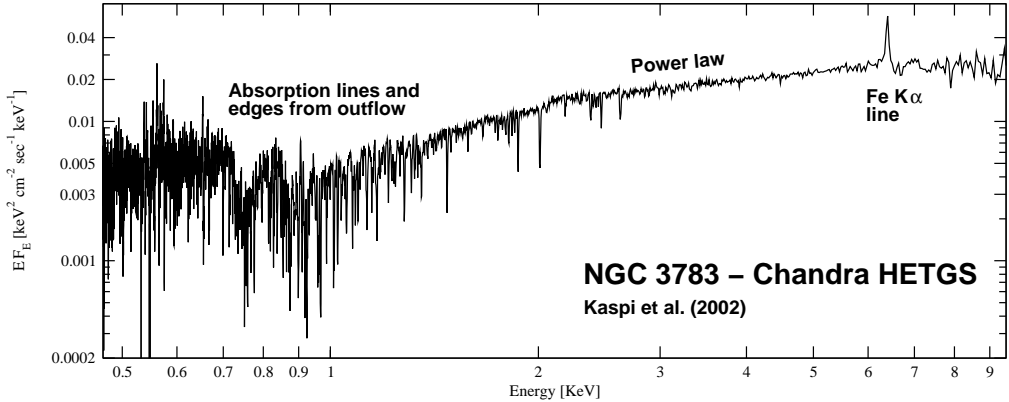}
\caption{\chandra\ High-Energy Transmission Grating Spectrometer
900~ks spectrum of the Seyfert~1 galaxy NGC~3783 illustrating absorption lines
and edges associated with an ionized outflow ($\approx 140$ such features are 
detected), the underlying power-law continuum, and the Fe~K$\alpha$ line.
The resolving power is $E/\Delta E\approx 250$--1500 over most of the plotted
spectral range. Adapted from Kaspi et~al. (2002).}
\end{figure}


\section{Bulk Follow-Up Work and AGN X-ray Evolution}

\subsection{Bulk SDSS Identification of X-ray Sources}

The SDSS spectroscopic survey is well matched in depth and sky density 
to the \rosat\ All-Sky Survey (RASS). Anderson et~al. (2003) describe 
plans to identify $\approx 10,000$ of the $\approx 100,000$ RASS sources
using the SDSS; objects in RASS error circles with unusual SDSS colors or 
FIRST radio detections are targeted. Thus far an impressive $\approx 1200$ AGN 
(about 80\% new) found in 1400~deg$^{2}$ have been delivered
($\approx 964$ broad-line AGN, 
$\approx 216$ intermediate AGN and Seyfert~2 candidates, and 
$\approx 45$ likely BL Lacs).
The ultimate order-of-magnitude increase in sample size compared to the 
largest previous X-ray source identification programs will allow
(1) studies of AGN X-ray evolution in the smallest possible luminosity
and redshift bins, and
(2) significantly improved statistical studies of minority X-ray AGN 
populations. Many bright ($g<17$) AGN are being discovered that deserve
targeted follow-up studies. 

\subsection{Bulk X-ray Investigation of SDSS Sources}

Vignali, Brandt, \& Schneider (2003a) have pursued a complementary
program: X-ray investigation of SDSS AGN serendipitously lying in 
pointed \rosat\ fields. In this case, the focus is on optically 
selected (rather than X-ray selected) AGN. This approach generates 
X-ray/SDSS matches not found by Anderson et~al. (2003) because
(1) the SDSS AGN density on the sky is several times higher than the 
RASS X-ray source density, and  
(2) pointed \rosat\ observations are typically 2--10 times 
more sensitive than the RASS. 
Vignali et~al. (2003a) find that $\approx 3.2$\% of the 462~deg$^2$ SDSS 
Early Data Release (EDR) area has sensitive coverage in pointed 
\rosat\ observations; this study presents pointed \rosat\ results for
128 of the EDR AGN in the Schneider et~al. (2002) catalog. This 
basic approach has large growth potential; ultimately $\approx 2700$
SDSS AGN are expected to have serendipitous pointed \rosat\ coverage. 

\subsection{AGN X-ray Evolution}

The bulk follow-up work summarized above enables multiple science 
projects, including investigations of the cosmic evolution of AGN 
X-ray emission. Such investigations can ultimately determine if AGN
black holes feed and grow in the same way at all cosmic epochs, 
and they have a long history going back to the 1980's (see
Vignali et~al. 2003a for a review). Generally, little X-ray evolution
is found with redshift after luminosity effects are taken into 
account, although there have been some notable counterclaims. 

Vignali et~al. (2003a) have recently used their SDSS AGN sample to
investigate the X-ray evolution of radio-quiet AGN. Their sample
has several notable advantages compared to previous work:
(1) it has been optically selected in a well-defined manner and spans
large ranges of redshift ($z=0.16$--6.28) and luminosity, 
(2) it has sensitive radio coverage from FIRST and NVSS enabling
effective selection of only radio-quiet AGN, and
(3) it allows the effects of Broad Absorption Line (BAL) quasars, which
often suffer from strong X-ray absorption, to be minimized and 
controlled. Partial correlation analyses using the parameter
\aox\ (the slope of a nominal power law between rest-frame 2500~\AA\ and 2~keV)
indicate no significant dependence of \aox\ upon redshift after
controlling for an observed dependence of \aox\ upon 
2500~\AA\ luminosity density ($l_{2500}$).
This strengthens and broadens earlier similar conclusions, 
demonstrating that the small-scale X-ray emission regions of most 
luminous AGN are remarkably stable from $z\approx 0$--6 despite 
the strong large-scale environmental changes over this interval. 

Anderson et~al. (2003) have investigated the dependence of \aox\ upon
$l_{2500}$ using their X-ray selected sample of
SDSS AGN. They suggest that the \hbox{\aox-$\log (l_{2500})$} 
relation may be nonlinear, with 
(1) relatively little dependence of \aox\ upon $l_{2500}$ 
for $l_{2500}\simlt 10^{29.5}$~erg~s$^{-1}$~Hz$^{-1}$, and 
(2) a stronger dependence of \aox\ upon $l_{2500}$ 
for $l_{2500}\simgt 10^{29.5}$~erg~s$^{-1}$~Hz$^{-1}$.
Similar evidence for nonlinearity has been presented by
Wilkes et~al. (1994) and Yuan et~al. (1998). 
This nonlinear dependence will need to be included in future 
studies of AGN X-ray evolution. 


\section{Selected Topics for Joint X-ray and SDSS Investigations}

\subsection{X-ray Weak AGN}

Although X-ray emission appears to be a universal property of AGN, there are
some notably X-ray weak AGN; these have $\alpha_{\rm ox}\simgt 1.75$--2 and lie on
a ``skew tail'' of the \aox\ distribution.\footnote{The appropriate \aox\
cutoff to use when defining an X-ray weak AGN is somewhat arbitrary, and 
the cutoff is likely luminosity dependent (see \S2.3). Aside from the
``skew tail'' toward large \aox, the \aox\ distribution for AGN of
similar luminosity appears to be roughly Gaussian. A reasonable 
criterion for X-ray weakness is that \aox\ should lie $>2\sigma$
from the mean of this Gaussian. Alternatively, mixture-modeling or
dip-test algorithms may be employed (see \S2 of BLW).}
X-ray weak AGN include even type~1 AGN where the view into the 
nucleus should not be blocked by the ``torus'' of AGN unification models. 
Such X-ray weakness could in principle be due to intrinsic X-ray 
absorption, a missing accretion-disk corona, or perhaps extreme X-ray 
variability. By determining the causes of \hbox{X-ray} weakness, we will refine 
our understanding of the universality of AGN X-ray emission. 

\begin{figure}[t!]
\plottwo{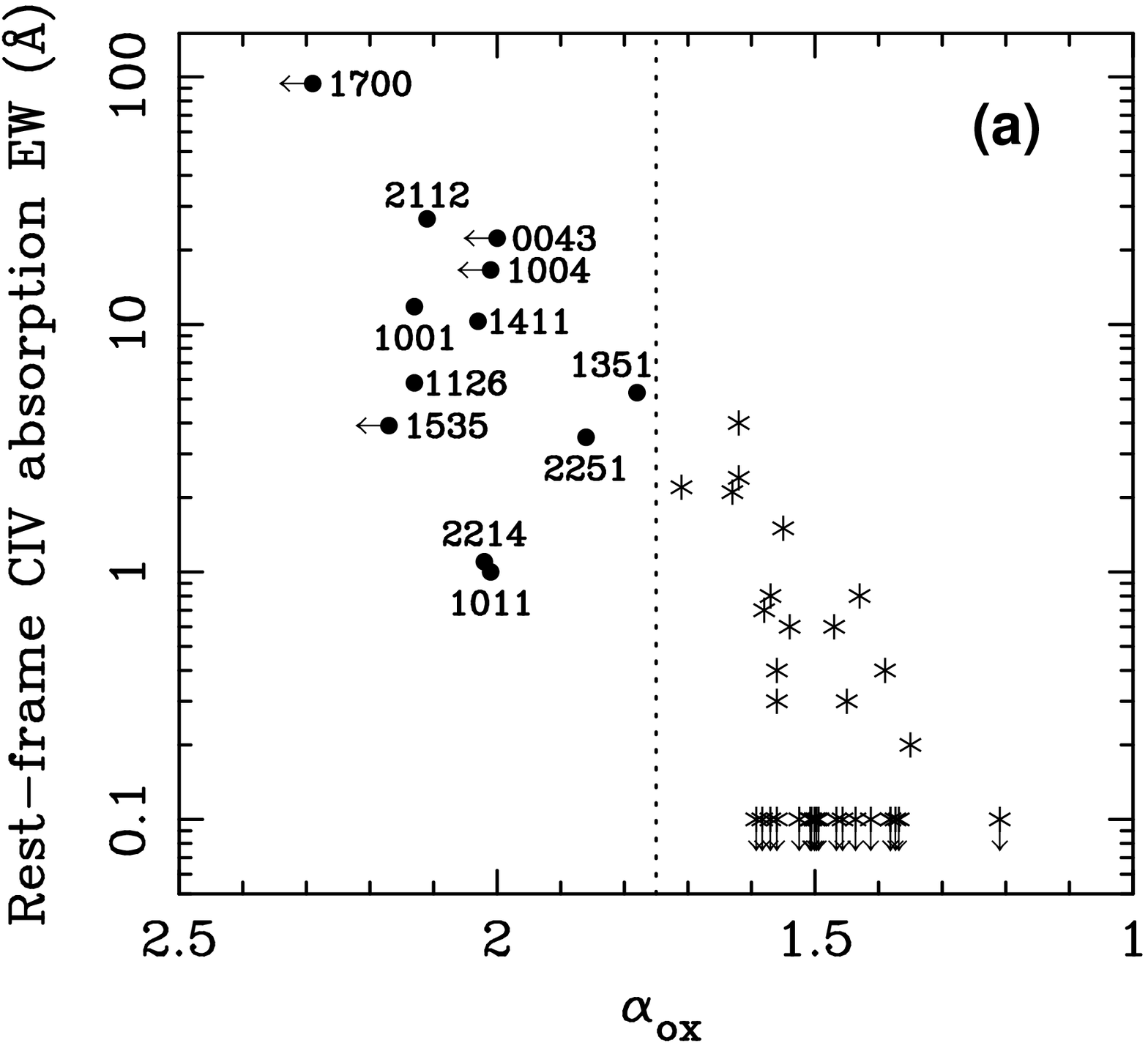}{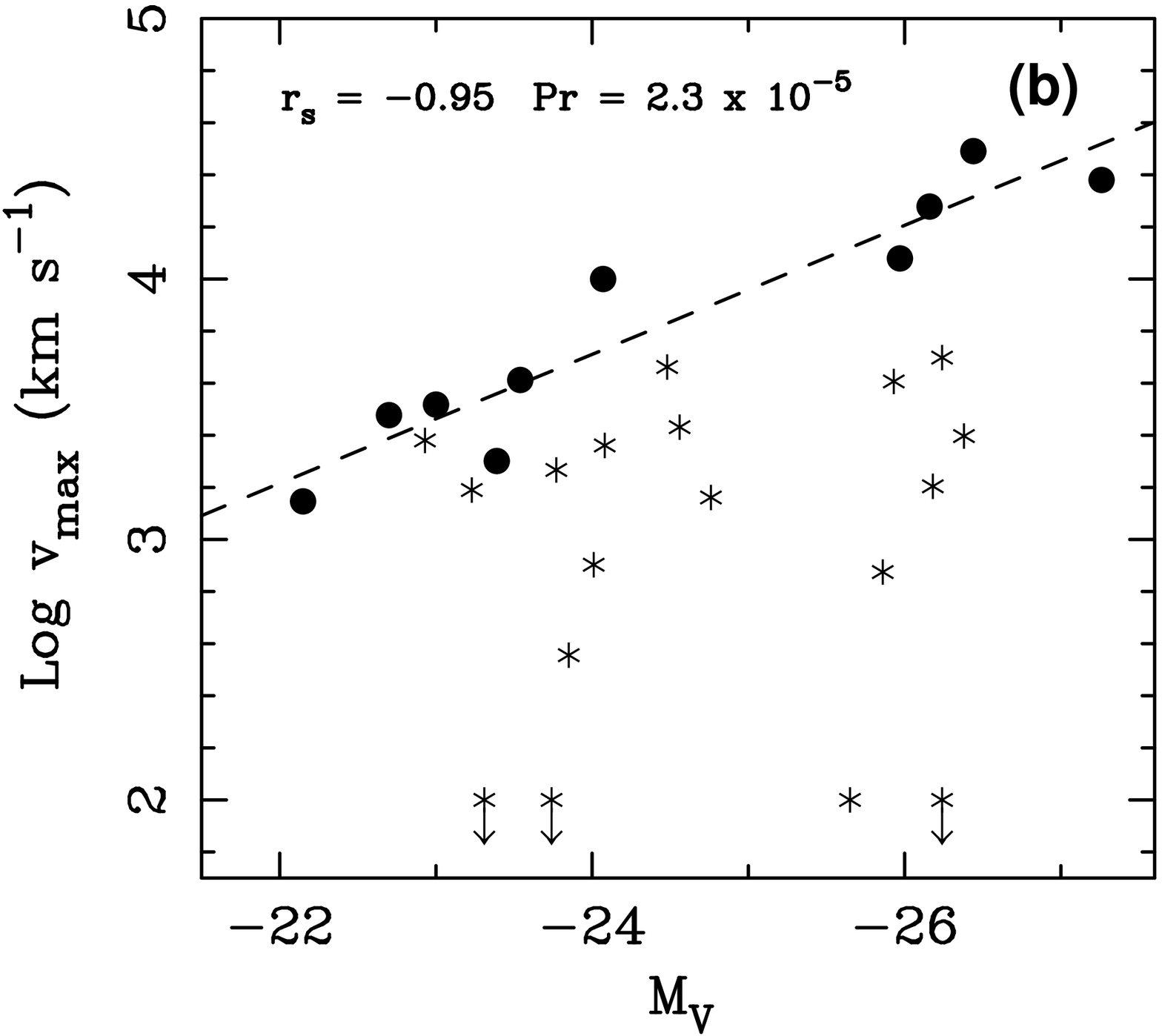}
\caption{(a) C~{\sc iv} $\lambda 1549$ absorption-line EW versus \aox\ for 
the \hbox{$z<0.5$} BQS AGN. X-ray weak AGN (filled circles 
with $\alpha_{\rm ox}>1.75$) are designated by the right ascension 
parts of their names. Stars denote objects with weak or no X-ray and UV absorption. 
Adapted from BLW. 
(b) The luminosity ($M_{\rm V}$) dependence of maximum UV absorption velocity
($v_{\rm max}$) for \hbox{$z<0.5$} BQS AGN. X-ray weak AGN (filled 
circles with $\alpha_{\rm ox}>2$) have the largest $v_{\rm max}$ at any given
luminosity. Stars denote BQS AGN with UV absorption that are not X-ray
weak. The dashed line shows the best-fit relation between $v_{\rm max}$
and luminosity: $v_{\rm max}\propto L^{0.62\pm 0.08}$ (see \S3.4
of Laor \& Brandt 2002 for comparison with simple radiation-pressure
driven outflow models). Adapted from Laor \& Brandt (2002).}
\end{figure}

Brandt, Laor, \& Wills (2000; hereafter BLW) systematically investigated X-ray 
weak AGN in the $z<0.5$ Bright Quasar Survey (BQS; Schmidt \& Green 1983). 
X-ray weak AGN comprise $\approx 11$--15\% of this 
blue AGN sample depending upon the (somewhat arbitrary)
\aox\ cutoff. From an observed strong correlation between \aox\ and the
equivalent width (EW) of blueshifted C~{\sc iv} absorption (see Fig.~2a), 
BLW argued that X-ray absorption associated with an outflow is likely the 
main cause of X-ray weakness for blue AGN.\footnote{This correlation is
also likely present for $z>4$ quasars; see \S4 of Vignali et~al. (2003c)
and references therein.}
This suspicion has been directly confirmed for several X-ray weak BQS AGN 
via X-ray spectroscopy (e.g., Brinkmann et~al. 1999; Gallagher et~al. 2001), 
although admittedly the spectroscopic sample size is limited. 

\begin{figure}[t!]
\plotone{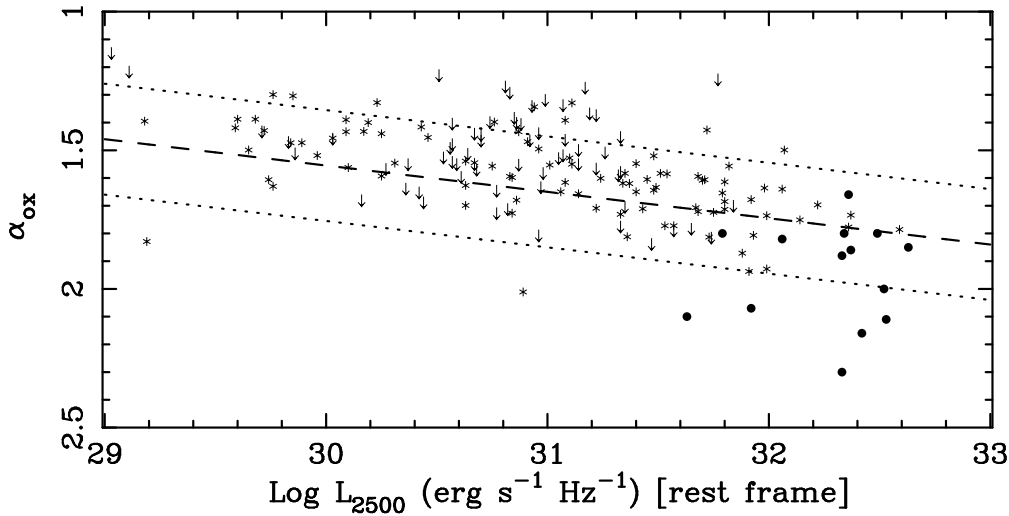}
\caption{\aox\ versus $l_{2500}$ for SDSS radio-quiet AGN (stars and 
downward-pointing arrows denoting detections and upper limits, respectively;
Vignali et~al. 2003a) and the HQS AGN put forward by Risaliti et~al. (2003)
as X-ray weak (filled circles). The dashed line represents the best fit
to the Vignali et~al. (2003a, 2003b) data using censored analysis
(see eqn.~4 of Vignali et~al. 2003b), and the dotted lines show $\pm 0.2$
intervals around the best fit.}
\end{figure}

It is also important to understand X-ray weakness in redder AGN; many
of these were missed by the BQS. Risaliti et~al. (2003) have recently 
investigated this issue using \chandra\ snapshot observations of 16 broad-line 
AGN from the grism-selected Hamburg Quasar Survey (HQS; e.g., Hagen et~al. 1995). 
These AGN are somewhat redder than BQS AGN, and 
they generally do not show evidence for X-ray
absorption. Risaliti et~al. (2003) propose that these AGN are intrinsically 
underluminous in X-rays and thus have intrinsic spectral energy distributions 
(SEDs) different from those of ``standard'' blue AGN. The overall situation 
would then be perplexing with
(1) X-ray weakness in blue AGN mainly being due to absorption, 
(2) X-ray weakness in somewhat redder AGN mainly being due to a different intrinsic SED, and 
(3) X-ray weakness in red AGN again mainly being due to absorption (see \S3.2). 
One possible resolution lies in the issue of sample definition. Specifically, 
most of the Risaliti et~al. (2003) AGN are highly luminous quasars, and 
many of these are only mildly X-ray weak compared to luminosity-dependent 
expectations for the range of \aox\ (see Fig.~3). That is, many do not lie 
on the ``skew tail'' (see above) of the \aox\ distribution but rather within 
the main body of this distribution. Thus, many of the Risaliti et~al. (2003) AGN 
are not directly comparable to, for example, the BQS AGN studied 
by BLW.\footnote{The application of a more stringent \aox\ criterion would
also significantly reduce the remarkably high fraction $(\simgt 50$\%)
of X-ray weak HQS AGN found
by Risaliti et~al. (2003).} The 4--5 Risaliti et~al. (2003) AGN 
that do appear to be comparably X-ray weak
have small numbers of counts ($<15$ to 53) in their \chandra\ snapshot 
observations, and the current X-ray spectral data do not strongly constrain 
potentially complex X-ray absorption in these objects. Further studies of 
these objects are needed. 

There are a few good candidates for intrinsically X-ray weak
AGN. One notable example is PHL~1811, which shows no evidence for 
absorption even in high-quality \chandra\ and \hst\ data
(Leighly et~al., these proceedings). Another, PG~1011--040, showed no 
evidence for X-ray absorption in an \asca\ spectrum (Gallagher et~al. 2001). 
The existence of these objects argues that X-ray absorption cannot 
universally explain X-ray weakness in AGN, and it is important to 
enlarge the sample of intrinsically X-ray weak AGN so that systematic 
studies are possible. 

The SDSS AGN sample combined with sensitive archival X-ray data should be able
to improve our understanding of X-ray weak AGN significantly. Ultimately a few
hundred X-ray weak SDSS AGN should be identified spanning broad ranges of
optical/near-infrared color, luminosity, and redshift, and follow-up studies of this large
sample should definitively determine the causes of X-ray weakness. Furthermore, 
it should be possible to quantify the luminosity dependence of AGN outflows
properly. Laor \& Brandt (2002) found that X-ray weak BQS AGN have the
highest maximum UV absorption velocity ($v_{\rm max}$) at any given 
luminosity ($M_{\rm V}$), and they suggested that $v_{\rm max}$ is largely set
by luminosity as expected for radiation-pressure driven outflows (see Fig.~2b). 
Refinement of such relations using SDSS AGN has the potential to provide a 
unified understanding of AGN outflows ranging from Seyfert ionized 
absorbers to BAL quasars. 

\subsection{Red AGN}

Moving to redder optical/near-infrared colors, significant X-ray attention
has recently been lavished upon 2MASS AGN with $J-K_{\rm s}>2$. These objects
appear to have a space density comparable to previously known blue AGN; about
80\% (20\%) are type~1 (type~2). Snapshot \chandra\ observations of $\approx 46$
of these objects reveal that they are generally X-ray weak (see Fig.~4a) and suggest
that their hard X-ray spectra are probably caused by intrinsic X-ray absorption
with $N_{\rm H}\approx 10^{21}$--$10^{23}$~cm$^{-2}$ (Wilkes et~al. 2002). There
appears to be little relation between the estimated amount of X-ray absorption
and near-infrared color, AGN classification, or optical polarization; hopefully
X-ray and UV spectroscopy will clarify the nature of the absorption. 

\begin{figure}[t!]
\plottwo{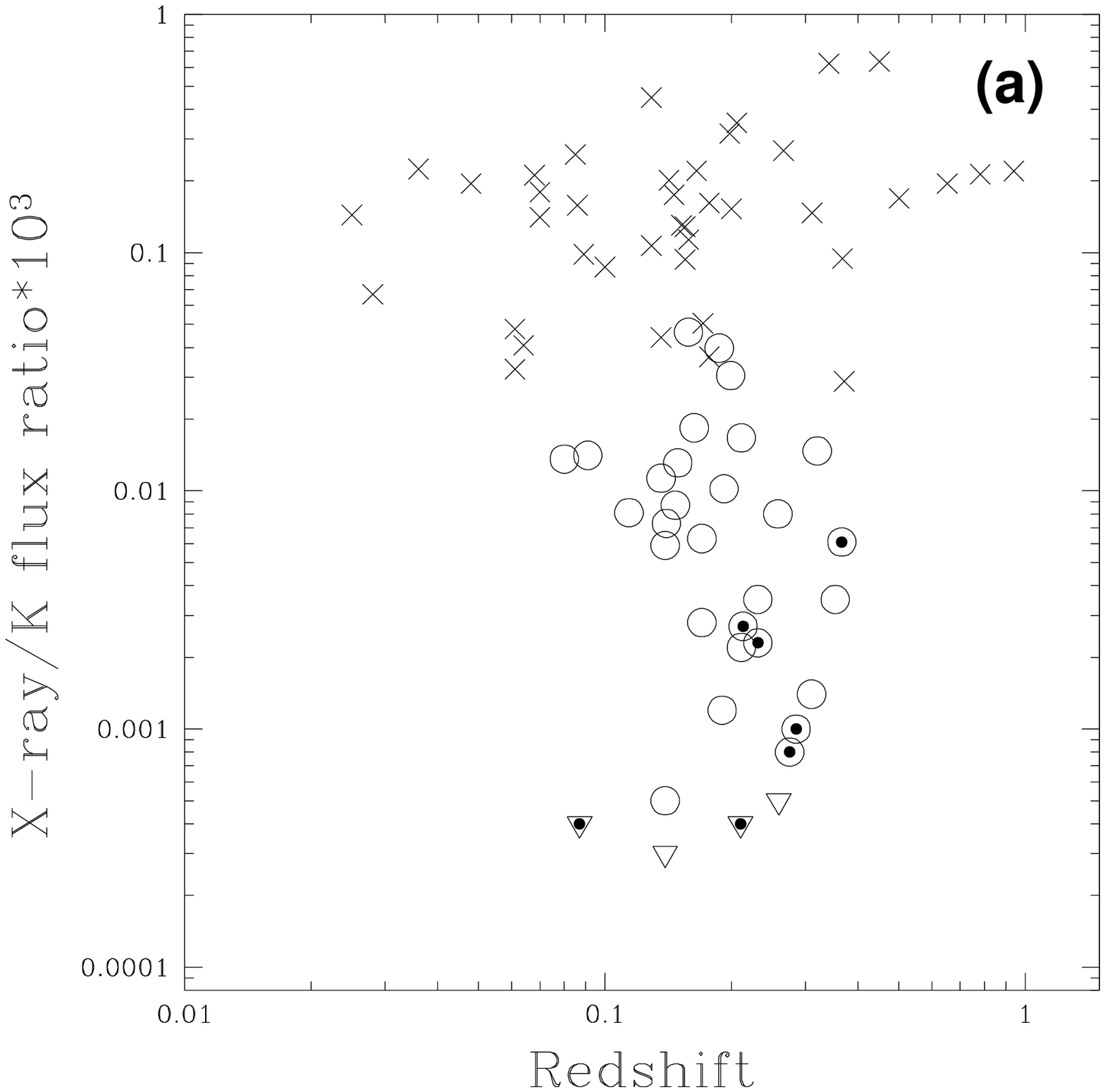}{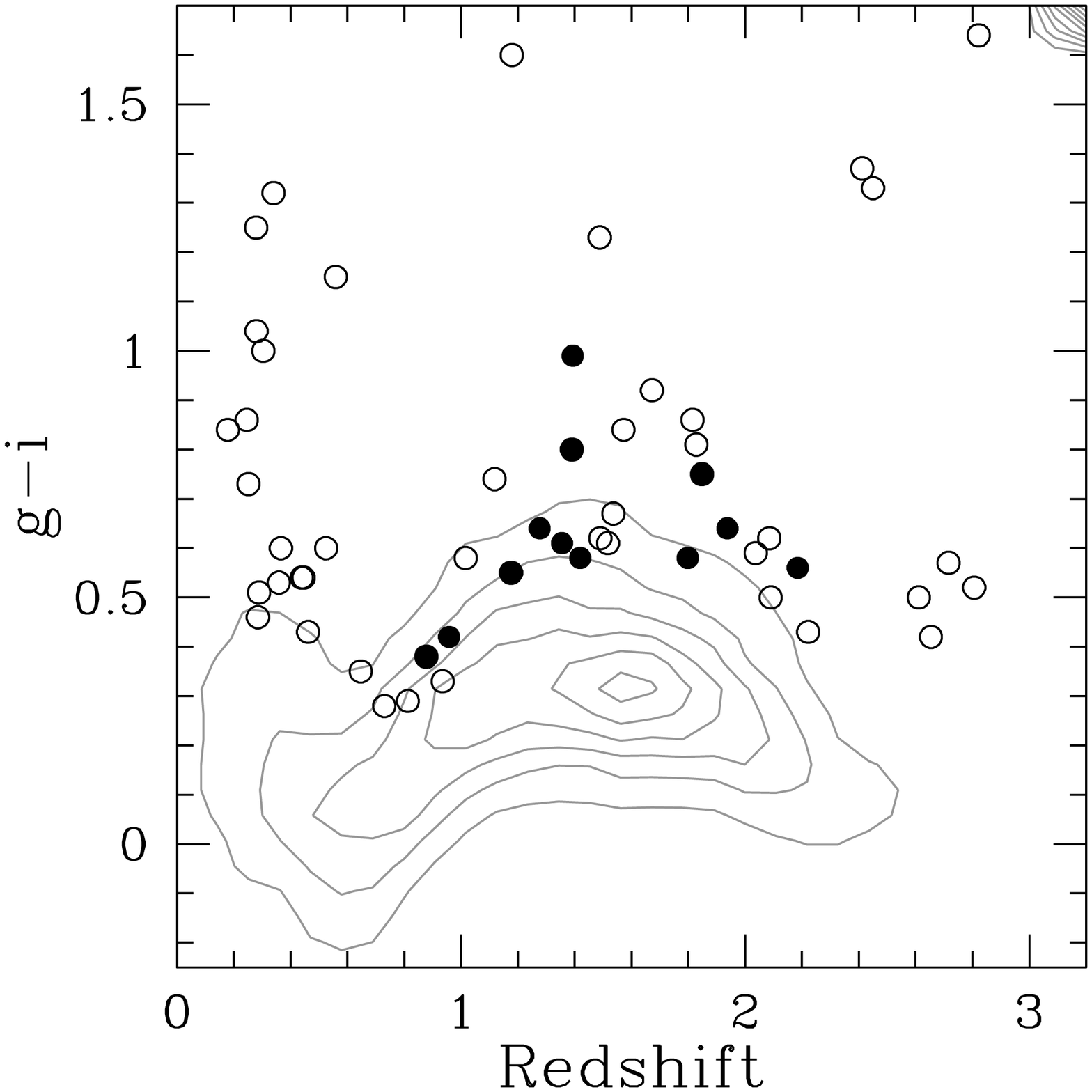}
\caption{(a) Observed X-ray (1~keV) to $K_{\rm s}$ flux ratio versus redshift
for 2MASS red AGN (circles) and low-redshift, broad-line AGN (crosses). Upper
limits are indicated by triangles, and the reddest sources with 
$J-K_{\rm s}>2.5$ are indicated by an enclosed square. From B.J. Wilkes (2003, 
private communication); updated from Wilkes et~al. (2002). 
(b) Observed $g-i$ color distribution of SDSS EDR AGN as a function of
redshift (contours). The circles are SDSS red AGN; filled circles
denote the 12 accepted for \chandra\ snapshot observations. Many of the
reddest AGN are BAL quasars and thus were not selected as 
\chandra\ targets. From G.T. Richards (2003, private communication).}
\end{figure}

Additional selection methods can broaden our understanding of red AGN
and allow X-ray studies out to high redshift; the Wilkes et~al. (2002) sample
has a small median redshift of $z\approx 0.2$. Richards et~al. (2003) have presented a 
complementary approach advocating the use of relative (rather than observed) SDSS
colors that remove redshift-dependent emission-line effects (see Fig.~4b). 
Their technique
(1) allows detection of more subtle reddening than for the 2MASS red AGN, 
(2) provides a more meaningful quantification of the amount of reddening, and
(3) allows high-redshift red AGN to be found effectively. \chandra\ snapshot 
observations have now been accepted for 12 of these SDSS red AGN 
at \hbox{$z=0.88$--2.19}, and comparisons with the 2MASS red AGN should 
be enlightening. 

\subsection{Hard X-ray Selected AGN}

Sensitive hard X-ray surveys reveal a wide variety of AGN. In
addition to the standard type~1 and type~2 AGN, a class of X-ray
bright, optically normal galaxies (XBONGs) is detected (e.g., 
Barger et~al. 2001; Hornschemeier et~al. 2001; Comastri et~al. 2002a,b). 
XBONGs can have hard X-ray spectra and X-ray luminosities
of $\approx 10^{41}$--$10^{43}$~erg~s$^{-1}$. Optical spectra
give redshifts of \hbox{$z\approx 0.05$--1}, but AGN emission lines
and non-stellar continua are not apparent. Some XBONGs may just
be Seyfert~2s where dilution by host-galaxy light hinders
optical detection of the AGN (e.g., Moran et~al. 2002; 
Severgnini et~al. 2003), but others have high-quality 
follow up and appear to be truly remarkable 
(e.g., Comastri et~al. 2002a). The place of the ``true'' XBONGs in the AGN 
unified model remains unclear; they may suffer from heavy obscuration 
covering a large solid angle ($\approx 4\pi$~sr) so that optical emission-line 
and ionizing photons cannot escape the nuclear region. XBONGs may be related
to local elusive AGN such as NGC~4945 and NGC~6240 
(e.g., Maiolino et~al. 2003). 

XBONGs are difficult to identify without X-ray data, and thus
they may elude selection in the SDSS AGN survey.
In fact, it was not possible to 
select the three XBONGs in the most intensively studied patch of 
sky, the Hubble Deep Field-North (HDF-N), prior to the \chandra\ 
observations (Brandt et~al. 2001); these three XBONGs are among 
the brightest X-ray sources in the HDF-N. 

Hundreds of archival hard X-ray observations combined 
with the SDSS should be powerful for finding local examples of 
XBONGs that can be studied in detail and hopefully placed within 
the unified model. Even if XBONGs are not selected in the SDSS 
AGN survey, $\simgt 200$ with sensitive hard X-ray coverage may 
be ``accidentally'' targeted for spectra in the SDSS galaxy 
survey. Joint hard X-ray and SDSS investigations can thus clarify 
both the poorly understood size of the ``true'' XBONG population and the 
overall AGN selection effectiveness of the SDSS. 

\subsection{High-Redshift AGN Demography}

\begin{figure}[t!]
\plottwo{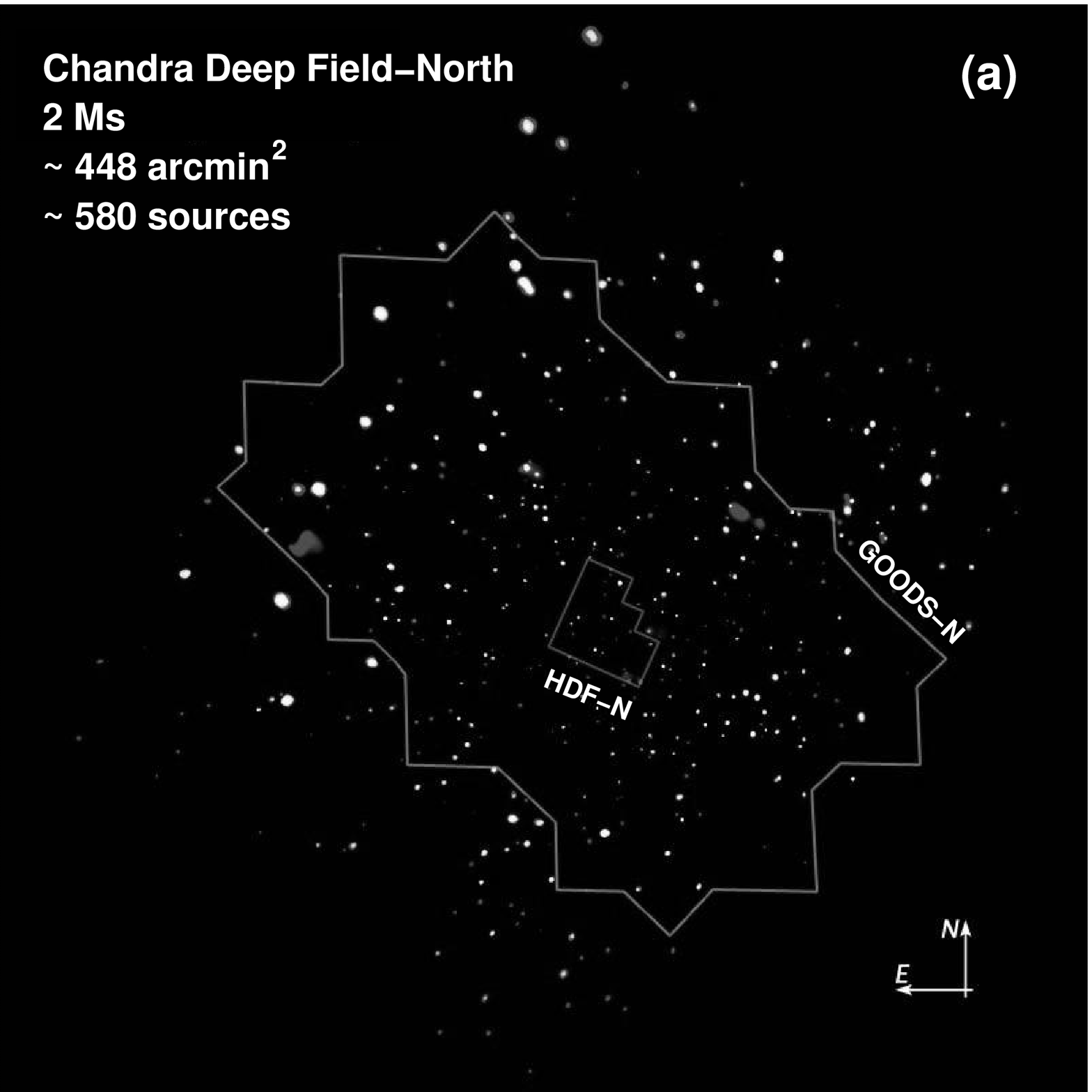}{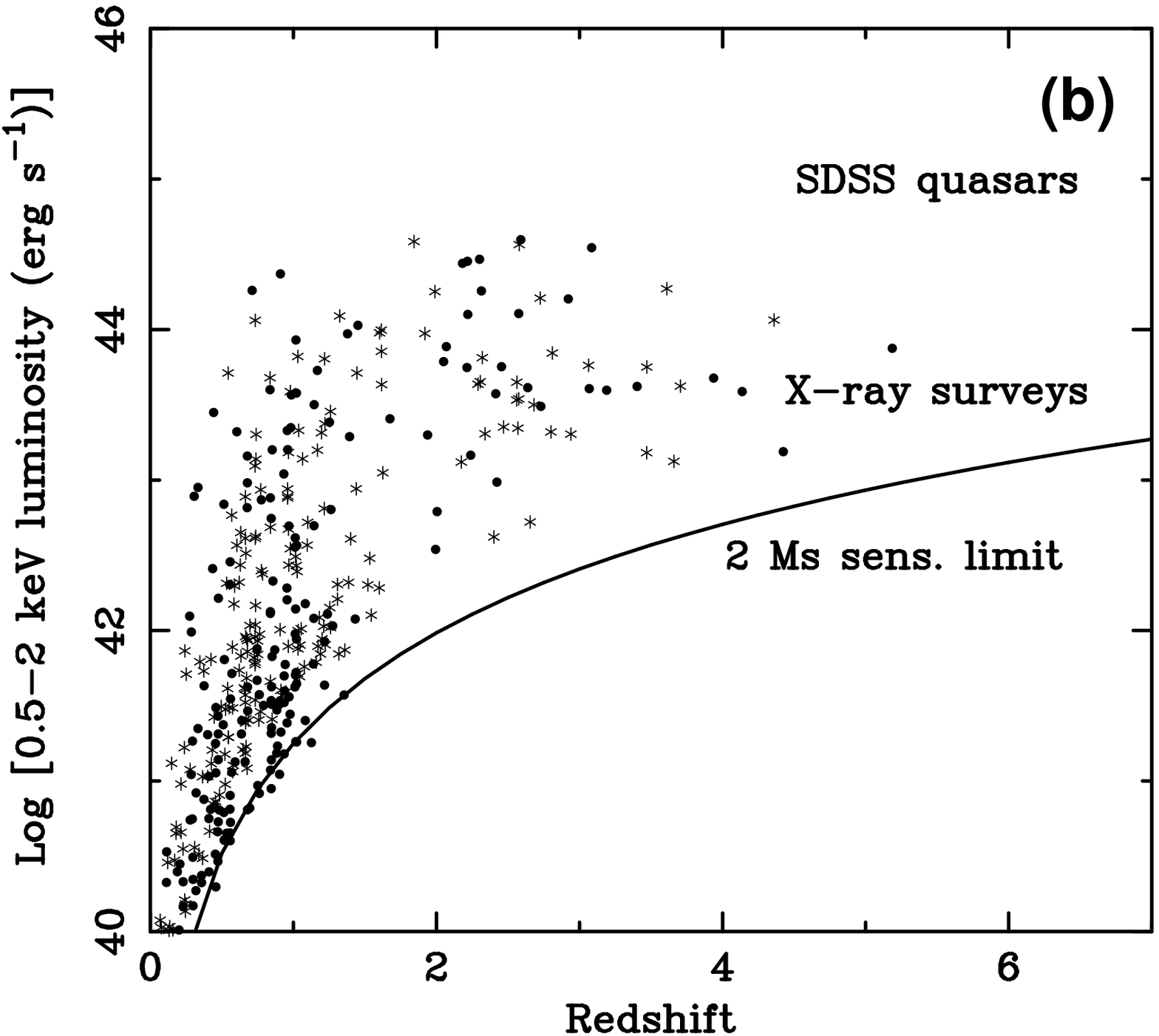}
\caption{(a) Adaptively smoothed image of the 2~Ms CDF-N. 
Nearly 600 sources are detected in the
$\approx 448$~arcmin$^2$ field ($\approx 1/80000$ the solid angle coverage
of the SDSS). The regions covered by the HDF-N and GOODS-N surveys 
are denoted. Adapted from Alexander et~al. (2003).  
(b) The $L_{\rm X}$--$z$ plane with identified AGN in the CDF-N 
and CDF-S plotted. The solid curve shows the 
on-axis 0.5--2~keV luminosity detection limit for 2~Ms. Note that
deep X-ray surveys can detect moderate-luminosity AGN out to $z>6$, 
while the SDSS only detects the most luminous quasars.}
\end{figure}

Some of the most exciting SDSS advances have been on luminous high-redshift
quasars (Fan et~al., these proceedings), and 
substantial effort has been devoted to X-ray follow-up
studies of these (e.g., Brandt et~al. 2003; Bassett et~al. and 
Vignali et~al., these proceedings). A key finding 
from this work is that, even at the 
highest redshifts, the basic \hbox{X-ray} emission properties of quasars do
not change strongly. This implies that X-ray AGN selection should remain
effective at high redshift, and it should be possible to combine the 
SDSS and X-ray surveys to explore high-redshift AGN demography. 
Sensitive X-ray surveys (e.g., Fig.~5a) can complement the SDSS since
they can detect AGN that are $\simgt 10$--30 times less luminous than the
SDSS quasars (see Fig.~5b); these lower luminosity AGN are much more 
numerous and thus more representative than the rare SDSS quasars. 
Furthermore, X-ray surveys suffer from minimal absorption bias at
high redshift where penetrating $\approx 2$--40~keV rest-frame X-rays 
are accessed. 

Follow-up studies of moderate-luminosity X-ray AGN at $z>4$ are challenging, 
but some progress has been made via large-telescope spectroscopy and
Lyman break selection (e.g., Barger et~al. 2003; Cristiani et~al. 2003; 
Koekemoer et~al. 2003). There are unlikely to be more than $\approx 8$ 
AGN at $z>4$ detectable in a $\approx 1$~Ms \chandra\ field, and the
combined SDSS and X-ray results indicate that the AGN contribution to
reionization at $z\approx 6$ is small. Additional deep ($\simgt 250$~ks)
\chandra\ observations are needed to provide more solid angle coverage
at sensitive flux levels, so that the AGN luminosity-function shape can
be determined at levels below those accessible to the SDSS; one such
project, the Extended \chandra\ Deep Field-South, has recently been
accepted for observation. Ultimately the SDSS, deep X-ray surveys, 
and wider X-ray surveys (e.g., ChaMP; see Silverman et~al. 2002) should be able 
to define the $z>4$ AGN luminosity function over a range of $\approx 1000$ 
in luminosity. 


\section{Future Prospects}

Future prospects for combined X-ray and SDSS studies appear wonderful! 
\chandra, \xmm, and the SDSS all continue to generate torrents of
superb data. \chandra\ and \xmm\ should each deliver
several thousand additional observations, and as the SDSS covers an
increasingly large solid angle the synergy with archival X-ray 
data becomes increasingly potent. The selected topics in \S2 and \S3
represent only a sampling of the science that should be enabled;
equally exciting research is expected on 
Narrow-Line Seyfert~1s, 
double-peaked line emitters, 
low-luminosity AGN, 
high-redshift quasars, 
type~2 quasars, and
BAL quasars 
(just to name a few examples). 

Future X-ray missions, some to be launched soon, will further
increase combined X-ray and SDSS science opportunities. The SDSS 
should be of utility, for example, in follow-up investigations
of many of the AGN detected in the $\approx 10$--150~keV \swift\ 
Burst Alert Telescope sky survey. \astroe2\ should allow 
high-quality iron~K$\alpha$ spectroscopy of selected X-ray
bright SDSS AGN. In the more distant future, X-ray missions
such as \conx, \xeus, and \genx\ should enable efficient 
high-quality X-ray spectroscopy for most SDSS AGN. 


\acknowledgments
We thank G. Risaliti and B.J. Wilkes for helpful discussions. 
Support from 
NASA LTSA grant NAG5-13035 (WNB, DPS), 
NSF CAREER award AST-9983783 (WNB), and
NSF grant AST-9900703 (DPS) 
is gratefully acknowledged. 


\end{document}